\documentclass[nohyper,letterpaper,notoc]{JHEP}

\usepackage{epsfig}

\def\Abar{\overline{A}}
\def\Bbar{\overline{B}}
\def\Cbar{\overline{C}}
\def\Dbar{\overline{D}}
\def\gsim{\ \rlap{\raise 3pt \hbox{$>$}}{\lower 3pt \hbox{$\sim$}}\ }
\def\lsim{\ \rlap{\raise 3pt \hbox{$<$}}{\lower 3pt \hbox{$\sim$}}\ }
\def\ifb{\,{\rm fb}^{-1}}
\def\iab{\,{\rm ab}^{-1}}
\def\pyr{{\rm yr}^{-1}}
\def\B0bar{\overline{B}{}^0}
\def\Lbabar{\mbox{{\LARGE B}\hspace{-0.1em}{\large A}\hspace{-0.1em}{\LARGE
B}\hspace{-0.1em}{\large A\hspace{-0.1em}R}}}
\def\babar{\mbox{B\hspace{-0.4em} {\scriptsize A}\hspace{-0.4em} B\hspace{-0.4em} {\scriptsize
A\hspace{-0.1em}R}}}
\title{\boldmath CP Tagged Decays at Super\Lbabar
\unboldmath}

\author{Adam F. Falk\\
Department of Physics and Astronomy,
The Johns Hopkins University\\
3400 North Charles Street, 
Baltimore, Maryland, U.S.A.\ 21218\\
\email{falk@jhu.edu}}

\abstract{We explore the possibility of measuring the CKM parameter $\gamma$ using CP tagged decays at a very high luminosity $e^+e^-$ $B$
Factory.  A new collider capable of integrating as much as $10\iab$ per year is
being discussed as a possible future for SLAC beyond the current PEP-II program, and could also be in the future of KEK.  In two years of operation, it could be possible for a successor to \babar\ or Belle to accumulate a sample of one million CP tagged $B_d$ decays.  We find that a theoretically clean extraction of $\gamma$ with uncertainty less than
$5^\circ$ may be achievable in the analysis of such a data set.
\\[0.8cm]
\centerline{\it(To appear in Physical Review D)}\\[0.6cm]
}

\preprint{JHU--TIPAC--2001-02\\
July 2001\\[0.15cm]
\hepph{0107066}}

\begin{document}

\section{Introduction}

The first years of running of the PEP-II and KEK-B $e^+e^-$ $B$ Factories
have been a remarkable success~\cite{BCP4talks}.  With instantaneous luminosities already
above design and a series of incremental upgrades planned,
it is reasonable to hope that $\babar$ and Belle will each accumulate as much as $500\ifb$ of integrated luminosity by 2006.  For these detectors to collect such enormous data sets 
would go an order of magnitude beyond what was anticipated when
they were designed and built~\cite{Harrison:1998yr,BelleTDR}.  Nevertheless,
further substantial upgrades to the SLAC $B$ Factory program are being considered.
The goals would be a collider capable of integrating as much luminosity as
$10\iab$ per year, and an upgraded detector, which might be called Super\babar, built to function in this new environment.  A very high luminosity $B$ Factory program could equally well be mounted at KEK as a successor to KEK-B and Belle.\footnote{A Hypernews Forum for discussion of detector issues and physics objectives for such a project based at SLAC is found at {\tt fermi.phy.uc.edu/HyperNews/get/forums/SuperBABAR.html}.  Because at the moment there is no analogous concrete discussion occurring for KEK , in what follows we will for simplicity call the proposed detector Super\babar, wherever it may eventually be built.}

An $e^+e^-$ collider running at $10\iab\pyr$ would produce 100 billion
$B\Bbar$ pairs per year.  The physics case for such an upgrade cannot
rest solely on the prospect of refining ongoing $B$ physics analyses with
another order of magnitude more data.  It is equally important to
identify new analyses which could only be done with such an enormous data
set.  Examples would include the studies of very rare decays such as
$B\to\tau\nu$, and of the detailed kinematics of rare decays such as $B\to
Ke^+e^-$.  Of particular interest are analyses which require the
restricted kinematics and overall quantum numbers implied by running at the $\Upsilon(4S)$,
since then there is no competition from BTeV or LHCb, both of which will
accumulate even larger raw data sets but in a messier and less constrained hadronic
environment.  It would be particularly valuable to identify a ``killer
application'' for Super\babar, an analysis so important and so unique to
this experiment that it could serve as the flagship justification for the
project.\footnote{It could be argued that the measurements of
$\sin2\alpha$ and $\sin2\beta$ were thought to be the ``killer apps'' for
the original $B$ Factories.  That the first turns out to be less
interesting than hoped and the second less unique to the $\Upsilon(4S)$ than
anticipated shows the dangers of relying too completely on such narrow
arguments, since the rich physics programs of \babar\ and
Belle clearly justify the experiments in any case.}

The purpose of this paper is to explore whether the clean measurement of
the CKM phase $\gamma$ from the analysis of CP tagged $B$ decays could serve this
purpose.  This method is certainly unique to the $\Upsilon(4S)$, and
it is certainly central to the $B$ physics program to measure $\gamma$
cleanly.  And it certainly does require collosal integrated luminosities. 
What we will find is that the analysis is somewhat on the edge of
feasibility even within a scenario of $10\iab\pyr$.  While the
extraction of $\gamma$ by this method has no theoretical uncertainties,
the accuracy which can be obtained depends on a number of as yet
unmeasured branching ratios and strong and weak phases.

After reviewing CP tagging in general as a tool for measuring $\gamma$ in
Section~\ref{sec:review}, we will go on in Section~\ref{sec:cptagging} to
discuss CP tagging in $B$ decays, where CP violation in $B^0-\B0bar$
mixing complicates the situation.  In Section~\ref{sec:gamma} we will
evaluate the extraction of $\gamma$ by this method at Super\babar\ in some
detail. Section~\ref{sec:conclude} offers some brief conclusions.

\section{CP tagging and the extraction of $\gamma$}\label{sec:review}

In the Wolfenstein parameterization~\cite{Wolfenstein:1983yz}, the two smallest elements of the CKM
matrix are the only ones with large CP violating phases:  $\arg
V_{ub}=-\gamma$ and $\arg V_{td}=-\beta$.  Because virtual top quarks
dominate $B^0-\B0bar$ mixing in the Standard Model and the quark
transition $b\to c\bar cs$ carries a negligible weak phase, the angle $\beta$ may be
measured cleanly from the time dependent CP asymmetry in $B\to\psi K_S$. 
There is no comparably straightforward method for measuring $\gamma$. 
Instead, a variety of complicated analyses have been proposed, each
with its own difficulties and uncertainties.  None of them is
certain as yet to lead to a clean and accurate measurement of $\gamma$.

The existing proposals to measure $\gamma$ fall into a number of
categories, distinguished by experimental requirements, theoretical
assumptions, and the precise parameters which are actually extracted
(typically the sine or cosine of $\gamma$, $2\beta+\gamma$ or
$2\beta+2\gamma$).  The Ur-method is to study CP violation in the
interference between mixing and decay in $B\to\pi^+\pi^-$, an
analysis which is sensitive to $\sin(2\beta+2\gamma)=\sin2\alpha$. 
Unfortunately, this method turns out not to be theoretically clean
because of significant penguin contributions to the decay amplitude.  The
solution to this problem is a more complicated isospin analysis, requiring
either measurements of $B\to\pi^0\pi^0$~\cite{Gronau:1990ka} or a study of the Dalitz plot for
$B\to\rho\pi$~\cite{Snyder:1993mx}.  The problem with $B\to\pi^0\pi^0$ is that this process is
rare and extremely hard to measure, while $B\to\rho\pi$ is not only
challenging experimentally but suffers from uncertainties in the
nonresonant contributions to $B\to3\pi$.  It is not at all clear that
either technique will succeed, either at the $\Upsilon(4S)$ or in a
hadronic $B$ experiment.

Another category of methods is to extract $\gamma$ directly from the rates
for the various rare two body decays
$B_{(u,d,s)}\to(\pi,K)$~\cite{Neubert:1998jq,Buras:1999rb}.  These analyses require
additional theoretical inputs, such as $SU(3)$ flavor symmetry and arguments for the
dynamical suppression of rescattering processes.  While the validity of
some assumptions can be checked in the data itself, these approaches leave
theoretical uncertainties which are hard to quantify reliably.  This limits
the precision with which they can be used to extract $\gamma$ by
themselves.  Nonetheless, they will provide important cross-checks on other
techniques, as well as help address the discrete ambiguities which theoretically cleaner methods leave unresolved.

In the long run, the most promising methods for measuring $\gamma$ involve
direct CP violation in $B$ and $B_s$ decays.  Time dependent studies of
the rare decays $B_s\to D^\pm K^\mp$ are sensitive to $\gamma$, but they
are accessible only to the hadronic experiments~\cite{Aleksan:1992nh}.  At the $\Upsilon(4S)$, one can extract $\gamma$ from triangle relations involving the decays $B^\pm\to(D^0,\Dbar^0) K^\pm$, where the $D^0$ and $\Dbar^0$ decay to a common final state such as $K^+\pi^-$~\cite{Atwood:1997ci}.  While this method is clean in principle, the combined branching ratios are of order $10^{-7}$ and the experimental feasibility depends on the triangles not being too ``squashed.''

In a recent paper, a new method for measuring $\gamma$ was proposed, based on triangle relations involving both flavor-tagged and ``CP tagged'' $B_s$ decays~\cite{Falk:2000ga}.  As an example, consider the final state $D_s^\pm K^\mp$.  Define the flavor tagged amplitudes
\begin{eqnarray} \label{ampl1}
  A_1 &=& A(B_s \to D_s^-K^+) \,,\nonumber \\
  A_2 &=& A(\Bbar_s \to D_s^-K^+) \,,\nonumber\\
  \Abar_1 &=& A(\Bbar_s \to D_s^+ K^-)\,,\nonumber \\
  \Abar_2 &=& A(B_s \to D_s^+ K^-)\,.
\end{eqnarray}
The magnitudes of these amplitudes may be measured at hadronic $B$ experiments, where the tagging is accomplished by the usual method of identifying the flavor of the bottom hadron on the other side of the event.  However, at an $e^+e^-$ $B$ Factory operating at the $\Upsilon(5S)$, one could go further.  Since in this environment the initial state is a CP eigenstate, one can tag the CP of a decaying $B_s$ if the decay on the other side is to a CP eigenstate such as $D_s^+D_s^-$.  Then CP tagged amplitudes may be defined, and they obey the triangle relations
\begin{eqnarray}\label{triangrels}
  A_{\rm CP} &=& A(B_s^{\rm CP}\to
  D_s^-K^+)=(A_1+A_2)/\sqrt2\,,\nonumber \\
  \Abar_{\rm CP} &=& A(B_s^{\rm CP} \to D_s^+
  K^-)=(\Abar_1+\Abar_2)/\sqrt2\,.
\end{eqnarray}
As written, the relations assume a particular phase convention for the CP transformation of the $|B_s\rangle$ state.  The extraction of $\gamma$ is independent of this convention; for a detailed discussion, see Ref.~\cite{Falk:2000ga}.

\FIGURE{\epsfig{file=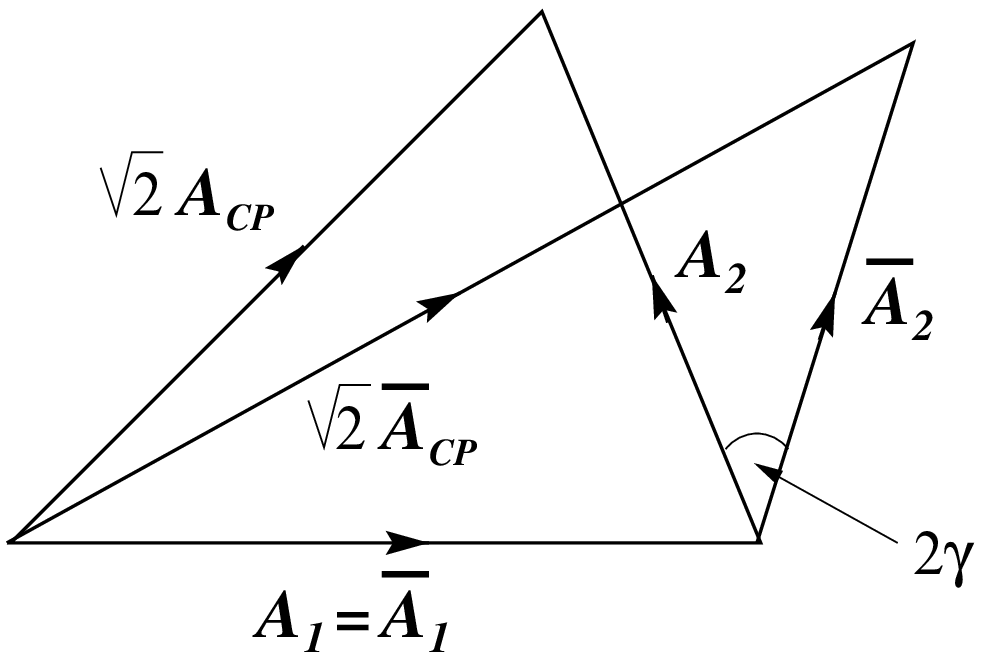,width=6cm}
\caption{\label{fig:triangle} Amplitude triangle relations for $B_s \to D_s K$.}}
The triangle relations are illustrated in Fig.~\ref{fig:triangle} in a phase convention for which $A_1=\Abar_1$.  Since $\Bbar_s\to D_s^-K^+$ is mediated by the quark transition $b\to u\bar cs$ which carries the weak phase $\gamma$, and $B_s\to D_s^+K^-$ is the CP conjugate process, it follows immediately that $2\gamma$ is the angle between the complex amplitudes $A_2$ and $\Abar_2$.   To derive an analytic expression for $\gamma$, we define the combinations of observables
\begin{eqnarray}\label{alphas}
  \alpha&=&{2|A_{\rm CP}|^2-|A_1|^2-|A_2|^2\over
  2|A_1||A_2|}\,,\nonumber\\
  \overline\alpha&=&{2|\Abar_{\rm CP}|^2-|\Abar_1|^2
  -|\Abar_2|^2\over2|\Abar_1||\Abar_2|}\,.
\end{eqnarray}
Then it follows that\footnote{Note that Eq.~(10) of Ref.~\cite{Falk:2000ga} is incorrect as written.}
\begin{equation}
  2\gamma=\arccos\alpha-\arccos\overline\alpha\,.
\end{equation}
There remains an eight-fold ambiguity in $\gamma$, taken over its full range $-\pi<\gamma\le\pi$.  With this ambiguity taken into account, decays which are tagged as CP even and CP odd contain the same information about~$\gamma$.

\section{CP tagging with CP violation in mixing}\label{sec:cptagging}

The simple method of tagging the CP of a decaying $B_s$ meson  by looking for a decay to a CP eigenstate on the other side is based on the Standard Model assumption that CP is conserved in $B_s$ mixing.  If a $B_s$ is known to be in a eigenstate of CP at time $t=0$, then when it decays at time $t$ its CP will be unaltered.  This assumption will be checked, since CP violation in $B_s$ will be constrained, and perhaps eventually measured, in future studies of asymmetries in the decay $B_s\to\psi\eta$.  The most serious difficulty with this method involves the fact that all the combinations $B_s\Bbar_s$, $B_s^*\Bbar_s$, $B_s\Bbar{}^*_s$ and $B_s^*\Bbar{}^*_s$ can be produced at the $\Upsilon(5S)$.  These samples must be separated for CP tagging.  It is also hard to see how one could accumulate sufficient statistics to make an accurate measurement feasible.

In the latter two respects, the situation is much better for $B_d$ meson pairs from $\Upsilon(4S)$ decay, since the cross section is larger and the $B_d^*$ is not produced.  On the other hand, it is complicated by the fact that in the Standard Model CP is violated in $B^0-\Bbar^0$ mixing.  The current world average $\sin2\beta=0.48\pm0.16$ implies that this effect has been confirmed at the $3\sigma$ level~\cite{sin2beta}.  To the extent that the amount of CP violation in mixing is known, it can be incorporated into the analysis.\footnote{Whether the value of ``$\sin2\beta$'' extracted from $B\to\psi K_S$ is of Standard Model origin is actually irrelevant here.  In the presence of contributions to $B^0-\B0bar$ mixing from new physics, ``$\sin2\beta$'' simply parameterizes the phase of the mixing amplitude.  When we write $\sin2\beta$ below, we really mean whatever parameter is extracted from the CP violating asymmetry in $B\to\psi K_S$.}  For a $B_d$ whose CP is tagged at $t=0$, with a perfect knowledge of $\sin2\beta$ and a perfect measurement of the decay time $t$, the admixture of CP even $B_d^+$ and CP odd $B_d^-$ states in the decaying $B$ meson would be known precisely as a function of $t$.  The evolution is given by
\begin{equation}
  B_d^\pm(t)=e^{-i(M_B+\Gamma/2)t}
  \left(a_\pm(t)B_d^\pm+b_\pm(t)B_d^\mp\right)\,,
\end{equation}
where
\begin{eqnarray}\label{evolution}
  a_\pm(t)&=&\cos\left({\Delta m_dt/2}\right)\pm
  i\cos2\beta\sin\left({\Delta m_dt/2}\right)\,,\nonumber\\
  b_\pm(t)&=&\mp\sin2\beta\sin\left({\Delta m_dt/2}\right)\,,
\end{eqnarray}
and $\Delta m_d$ is the mass splitting between $B_H$ and $B_L$.

With enough statistics, one could analyze the triangle relations for a fixed value of $t$ and extract $\gamma$ as cleanly as before.  Such a method is outlined in Ref.~\cite{Falk:2000ga}, and we recall the results here.  For pedagogical reasons, let us discuss the decay $B_d\to D^\pm\pi^\mp$.  In this case the amplitude triangles are actually ``squashed'' and the experiment is unlikely to be feasible.  However, it is illustrative because it depends on the transitions $b\to c\bar ud$ and $b\to u\bar cd$, for which it is straightforward to follow the flavor of the light quarks. 

Defining the amplitudes
\begin{eqnarray}
  A_1 &=& A(B_d\to D^-\pi^+) \,,\nonumber \\
  A_2 &=& A(\Bbar_d \to D^-\pi^+) \,,\nonumber\\
  \Abar_1 &=& A(\Bbar_d \to D^+ \pi^-)\,,\nonumber \\
  \Abar_2 &=& A(B_d \to D^+ \pi^-)\,,\nonumber\\
  A_\pm(t) &=& A(B_d^\pm(t) \to D^-\pi^+)\,,
  \nonumber \\
  \Abar_\pm(t) &=& A(B_d^\pm(t) \to D^+\pi^-)\,,
\end{eqnarray}
the conclusion of Ref.~\cite{Falk:2000ga} is that the expressions for $\alpha$ and $\overline\alpha$ become time dependent,
\begin{eqnarray}\label{alphast}
  \alpha_\pm(t)&=&{2|A_\pm(t)|^2
  -r_\mp^2(t)|A_1|^2-r_\pm^2(t)|A_2|^2
  \over2r_+(t)r_-(t)|A_1||A_2|}\,,\nonumber\\
  \overline\alpha_\pm(t)&=&{2|\Abar_\pm(t)|^2
  -r_\pm^2(t)|\Abar_1|^2-r_\mp^2(t)|\Abar_2|^2
  \over2r_+(t)r_-(t)|\Abar_1||\Abar_2|}\,,
\end{eqnarray}
where
\begin{equation}
  r_\pm(t)=\big[1\pm\sin2\beta\sin\Delta m_dt\big]^{1/2}.
\end{equation}
The cases in which the $B$ mesons are tagged at $t=0$ as CP even and CP odd are not equivalent until one extracts $\gamma$ via
\begin{equation}
  2\gamma=\arccos\alpha_\pm(t)-\arccos\overline\alpha_\pm(t)\,,
\end{equation}
at which point the time dependence cancels as well.  What happens is that the amplitude triangles such as those in Fig.~\ref{fig:triangle} change with $t$, but the angle $\gamma$ does not.

Accumulating enough statistics to perform the triangle analysis in a single individual bin in $t$ is far beyond the reach of any current or planned $B$ Factory.  Although a simultaneous fit in many $t$ bins could be performed, all restricted to a common value of $\gamma$, such an analysis would be complicated, with error depending in an intricate way on the experimental resolution in $t$ and the uncertainty in $\sin2\beta$.  While this may eventually prove practical, it would clearly be preferable to extract $\gamma$ from time integrated rates for which the explicit binning in $t$ did not have to be performed.  We now outline such a method.  Although there will be no direct graphical interpretation, the analytical result will be quite similar to the previous case.

The flavor tagged amplitudes may be parameterized in terms of their magnitudes and strong and weak phases,
\begin{eqnarray}\label{ampdefs}
  &&A_1=a_1 e^{i\delta_1}\,,\qquad\qquad\qquad \Abar_1=a_1 e^{i\delta_1}e^{-2i\xi}e^{2i\eta}\,, 
  \nonumber\\
  &&A_2=a_2 e^{-i\gamma}e^{i\delta_2}\,,\qquad\qquad\,\,
   \Abar_2=a_2 e^{i\gamma}e^{i\delta_2}e^{2i\xi}e^{2i\eta}\,,
\end{eqnarray}
where $\delta_1$ and $\delta_2$ are CP conserving strong phases, and $\xi$ and $\eta$ are the arbitrary phases associated with CP transformations of the initial and final states, $CP|B_d\rangle=e^{2i\xi}|\Bbar_d\rangle$ and $CP|D^-\pi^+\rangle=e^{2i\eta}|D^+\pi^-\rangle$.  The triangle relations are then
\begin{eqnarray}
  \sqrt2\,A_\pm(0)&=&A_1\pm e^{2i\xi}A_2=e^{i\delta_1}(a_1\pm a_2e^{-i\gamma}e^{i\delta})\,,
  \nonumber\\
  \sqrt2\,\Abar_\pm(0)&=&\Abar_2\pm e^{2i\xi}\Abar_1=\pm e^{i\delta_1}e^{2i\eta}
  (a_1\pm a_2e^{i\gamma}e^{i\delta})\,,
\end{eqnarray}
where $\delta=\delta_2-\delta_1+2\xi$ is a physical parameter.  In terms of the time dependent amplitudes, we find
\begin{eqnarray}\label{triangrels2}
  \sqrt2\, A_\pm(t)&=&e^{i\delta_1}\left[\left(a_\pm(t)+b_\pm(t)\right)a_1
  \pm\left(a_\pm(t)-b_\pm(t)\right)a_2e^{-i\gamma}e^{i\delta}\right]\,,\nonumber\\
  \sqrt2\, \Abar_\pm(t)&=&\pm e^{i(\delta_1+2\eta)}\left[\left(a_\pm(t)+b_\pm(t)\right)a_1
  \pm\left(a_\pm(t)-b_\pm(t)\right)a_2e^{i\gamma}e^{i\delta}\right]\,.
\end{eqnarray}
Note that $a_\pm(t)$ and $b_\pm(t)$ are themselves complex.

For the CP tagged decays, it is convenient to work in terms of the observable quantities
\begin{equation}
  C_\pm={\Gamma\over2}\int_{-\infty}^\infty {\rm d}t\,|A_\pm(t)|^2\qquad {\rm and}\qquad
  \Cbar_\pm={\Gamma\over2}\int_{-\infty}^\infty {\rm d}t\,|\Abar_\pm(t)|^2\,,
\end{equation}
which are respectively the time averaged branching ratios for the processes $B_d^\pm(t)\to D^-\pi^+$ and $B_d^\pm(t)\to D^+\pi^-$.  Recall that at the $\Upsilon(4S)$ the tag can come either before or after the decay.  It is then straightforward to compute
\begin{eqnarray}
  2C_\pm&=&a_1^2+a_2^2\pm2a_1a_2g_-\,,\nonumber\\
  2\Cbar_\pm&=&a_1^2+a_2^2\pm2a_1a_2g_+\,,
\end{eqnarray}
where
\begin{equation}
  g_\pm=\cos(\delta\pm\gamma)(1-2\chi_d\sin^22\beta)
 +2\sin(\delta\pm\gamma)\chi_d\sin2\beta\cos2\beta\,.
\end{equation}
The parameter $\chi_d=(\Delta m_d)^2/2[(\Delta m_d)^2+\Gamma^2]=0.174\pm0.009$ is known accurately from $B_d$ mixing measurements~\cite{Groom:2000in}.  The amplitudes $a_i$ can be extracted from flavor tagged branching ratios, which are much easier to measure than the CP tagged ones.

The next step is to eliminate $\delta$ and extract $\gamma$ in terms of the measured branching fractions and the mixing parameters.  This is accomplished most easily by writing the expressions in terms of $g_+\pm g_-$, which leads to
\begin{eqnarray}
  &&{C_\pm-\Cbar_\pm\over2a_1a_2}=\pm\sin\gamma\left[(1-2\chi_d\sin^2 2\beta)\cdot\sin\delta
  -2\chi_d\sin2\beta\cos2\beta\cdot\cos\delta\right]\,,\\
  &&{C_\pm+\Cbar_\pm-a_1^2-a_2^2\over2a_1a_2}=
  \pm\cos\gamma\left[(1-2\chi_d\sin^2 2\beta)\cdot\cos\delta
  +2\chi_d\sin2\beta\cos2\beta\cdot\sin\delta\right]\,.\nonumber
\end{eqnarray}
Defining an angle $\alpha_1$ by
\begin{equation}
  \cot\alpha_1=(1-2\chi_d\sin^2 2\beta)/2\chi_d\sin2\beta\cos2\beta\,,
\end{equation}
we obtain simple relations in terms of newly defined observables $x$ and $y$,
\begin{eqnarray}\label{xydef}
  &&x\equiv\pm{C_\pm-\Cbar_\pm\over2Ra_1a_2}=\sin\gamma\sin(\delta-\alpha_1)\,,\nonumber\\
  &&y\equiv\pm{C_\pm+\Cbar_\pm-a_1^2-a_2^2\over2Ra_1a_2}=\cos\gamma\cos(\delta-\alpha_1)\,,
\end{eqnarray}
where the dilution factor $R$ is given by
\begin{equation}
  R=\left[(1-2\chi_d\sin^2 2\beta)^2+(2\chi_d\sin2\beta\cos2\beta)^2\right]^{1/2}\,.
\end{equation}
Note that the dependence on the angle $\alpha_1$ is such that it will be eliminated together with  the strong phase difference $\delta$.  These equations are analogous to the sum and difference of the expressions in Eq.~(\ref{alphas}), and we find a simple result for $\gamma$,
\begin{equation}\label{gamsolution}
  2\gamma=\arccos(x+y)-\arccos(x-y)\,.
\end{equation}
There is an eightfold ambiguity in $\gamma$, since each $\arccos$ is double valued and the expression for $2\gamma$ must be mapped into $-\pi\le\gamma<\pi$.  If we take $\gamma_*$ to be the solution in the range $0\le\gamma_*\le\pi/4$, then the eight possible values of $\gamma$ are $\{\pm\gamma_*,\pm({\pi/2}-\gamma_*),\pm({\pi/2}+\gamma_*),\pm(\pi-\gamma_*)\}$.  If $\gamma_*=0$ or ${\pi/4}$, then the solutions are pairwise degenerate.  Because $\arccos(-z)=\pi-\arccos z$, the $\pm$ factors in Eq.~(\ref{xydef}) are absorbed in Eq.~(\ref{gamsolution}) into the sign of $\gamma$, which this method cannot determine.  So in view of the ambiguities, it makes no difference whether the decay is tagged as CP even or CP odd.  To emphasize this, w give the time averaged branching ratios $C_\pm$ and $\Cbar_\pm$ the new names $C_{\rm CP}$ and $\Cbar_{\rm CP}$.  Then we find the equivalent result
\begin{equation}
  2\gamma=\arccos\left[{2C_{\rm CP}-a_1^2-a_2^2\over2Ra_1a_2}\right]
  -\arccos\left[{2\Cbar_{\rm CP}-a_1^2-a_2^2\over2Ra_1a_2}\right]\,,
\end{equation}
in complete analogy to Eq.~(\ref{alphas}).  As it turns out, the dilution factor $R$ is the only new element which is introduced by CP violation in $B_d$ mixing.  In the limits $\chi_d\to0$ and $\sin2\beta\to0$ the CP violating effects vanish, and $R\to1$ as one would expect.  With $\chi_d=0.174$ and taking the world average value $\sin2\beta=0.48\pm0.16$, we find $0.87<R<0.97$ and $6.2^\circ<\alpha_1<11.3^\circ$.  If $\sin2\beta$ is unrestricted, then $0.65<R<1$ and $\alpha_1<12.2^\circ$.

As we noted before, the amplitude triangles for $B_d\to D^\pm\pi^\mp$ are ``squashed'', since the amplitude for $b\to u\bar cd$ is suppressed relative to that for $b\to c\bar ud$ by a factor of order $\sin^2\theta_C$.  This makes the analysis impractical for these modes.  A preferable alternative is to use the modes $B_d\to(D^0,\Dbar{}^0)K_S$ or $B_d\to(D^0,\Dbar{}^0)K_L$, which are mediated by the quark transitions $b\to u\bar cs$ and $b\to c\bar us$, both of order $\sin^3\theta_C$.  The CP of the neutral kaon state affects the ampliudes only through the phase $\eta$, which as we see from Eq.~(\ref{triangrels2}) cancels in the observable rates.

\section{Extraction of $\gamma$ with CP tagged decays at Super\babar}\label{sec:gamma}

We turn now to the question of the accuracy in $\gamma$ which one could expect to achieve with this method, if one had at one's disposal an $e^+e^-$ collider running at the $\Upsilon(4S)$ and integrating $10\iab$ per year.  To be concrete, we will consider the $B_d\to DK_S$ channel.  Of course, it remains unclear whether such an accelerator and its associated detector can or will ever be built, and it is not our purpose here to advocate this project.  For the moment, we assume only that any such machine would take data no earlier than the end of the decade, after the current $B$ Factory programs and concurrent with or after the start of BTeV and LHCb.  By that time, we expect that the relevant flavor tagged branching ratios will have been measured accurately elsewhere, as will have $\sin2\beta$.  The only significant source of experimental uncertainty will be the CP tagged branching ratios, which it would be up to Super\babar\ to determine.  Note that although a next generation $B$ Factory may well be asymmetric, this feature is not needed to determine the time integrated CP tagged branching ratios $C_{\rm CP}$ and $\Cbar_{\rm CP}$.

The most practical mode for CP tagging will be $B_d\to\psi K_S$, because of its relatively large branching ratio and clean experimental signature.  The $\psi K_L$ channel may also be a possibility, although its reconstruction efficiency is lower.  According to Ref.~\cite{Harrison:1998yr}, the \babar\ detector would fully reconstruct a sample of 770 $B^0\to\psi K_S$ decays from a data set corresponding to $30\ifb$ of integrated luminosity.  Recognizing that the Super\babar\ detector would in fact be quite different from \babar, let us nevertheless scale na\"\i vely to a two year run of a next generation $B$ Factory, corresponding to an integrated luminosity of $20\iab$.  Since either side of the event can be tagged, this would correspond to a sample of $10^6$ CP tagged $B_d$ mesons whose decays could then be studied.

Since they have not been measured, we need to estimate the branching fractions relevant to our analysis.  Given that ${\cal B}(B^-\to D^0K^-)\approx3\times10^{-4}$~\cite{Groom:2000in}, we take ${\cal B}(B_d\to DK_S)\approx{\cal B}(B_d\to D^*K_S)\approx3\times 10^{-4}$ as well.  If the efficiency for reconstructing these final states were 15\%, then it would be possible to accumulate approximately 50 events per channel.  This would be enough for a measurement of the CP tagged branching ratio with a statistical error of 15\% in each mode.  

\TABLE{
  \begin{tabular}{|c|c|}
  \hline  Parameter & Values\\ \hline
  \hline  $\gamma$ & $70^\circ,\quad 110^\circ$\\
  \hline  $\sin2\beta$ & $0.32,\quad 0.48,\quad 0.64$\\
  \hline  $a_2/a_1$ & $1,\quad2$\\
  \hline  $\bar\delta\equiv\delta-\alpha_1$ & $10^\circ,\quad 60^\circ$\\
  \hline  $\Delta$ & $10\%,\quad 20\%$\\
  \hline
  \end{tabular}\caption{Parameters for the extraction of $\gamma$.\label{table:params}}}

What accuracy on $\gamma$ would such a measurement imply?  In principle, the accuracy of the determination depends on quantities such as $a_2/a_1$, $\delta$, and $\sin2\beta$, as well as the actual value of $\gamma$ itself.  In view of this, we examine the extraction of $\gamma$ under the various assumptions listed in Table~\ref{table:params}.  The values of $\gamma$ are representative ones which roughly bracket the Standard Model expectation~\cite{Hocker:2001xe}.  For $\sin2\beta$, we take the one standard deviation about the world average  Since $a_1$ and $a_2$ play interchangeable roles in the analysis, we consider the $a_2=a_1$ case and a single reasonable $a_2>a_1$ case.   It is certainly possible that $a_2/a_1$ (or $a_1/a_2$) is larger than this, but we know the method will fail if the two amplitudes are not roughly comparable.  In any case, the ratio will be known from the flavor tagged branching ratios before this experiment is performed.  The parameter $\bar\delta$ turns out to be important, and we take two values, one large and one small.  Since $\alpha$ is of order $5^\circ-10^\circ$, it would be unnatural for $\bar\delta$ to be much smaller than this.  However, it is also unlikely for it to be much larger, since $\delta$ vanishes in the flavor SU(3) limit.  In Ref.~\cite{Falk:2000ga} it was argued that in $B_s\to D_sK$ the analogous phase is probably less than $5^\circ$.  That numerical estimate depended on a model based on Regge phenomenology~\cite{Falk:1998wc}, but the argument really is more general;  it is difficult to find explicit SU(3) violating effects which would make $\delta$ large.  Here we include the possibility $\bar\delta=60^\circ$ for completeness and to explore its implications, but in fact we think that such a large value is not reasonable.  Finally, we denote by $\Delta$ the precision with which the CP tagged branching fractions are measured.  To explore the impact of $\Delta$ on the extraction of $\gamma$, we will consider $\Delta=10\%$ and $20\%$, values which bracket our rough expectation of $\Delta=15\%$.

Our procedure is to use a set of input parameters to predict the observables, then extract $\gamma$ from the ``data'' under the assumption that the only significant experimental uncertainty is in $C_{\rm CP}$ and $\Cbar_{\rm CP}$.  The result is eight allowed regions for $\gamma$, possibly overlapping, one of which contains the ``true'' value.  In the plots to follow, the allowed regions are shown as grey bands, while the true value of $\gamma$ is indicated by a solid line.  We only display the solutions with $\gamma\ge0$, leaving the $\gamma\to-\gamma$ ambiguity implicit.

\FIGURE{ \epsfig{file=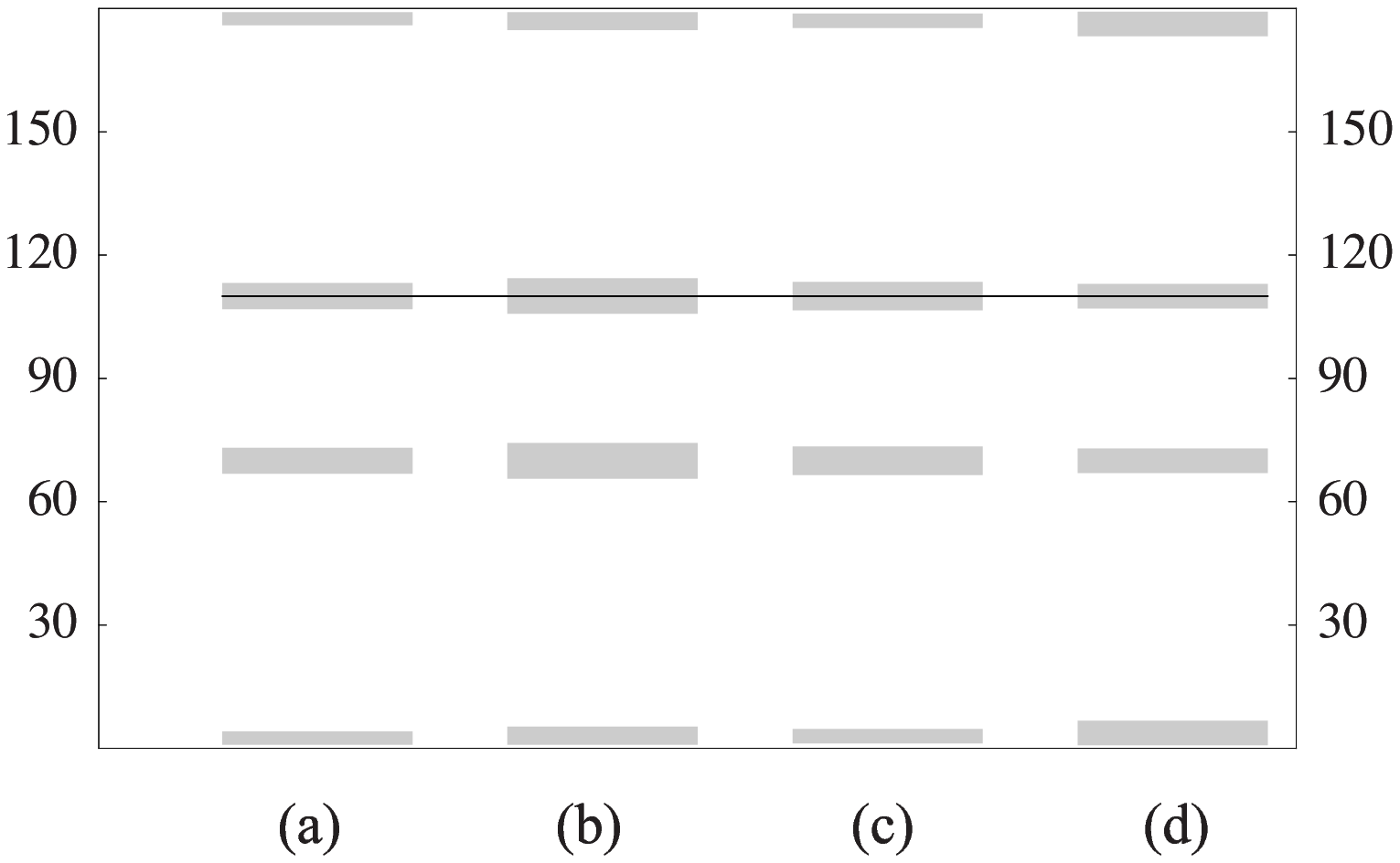,width=12cm}
\caption{\label{fig:gamplot1} Allowed bands for $\gamma$, with input values $\gamma=110^\circ$, $\Delta=10\%$, $\bar\delta=10^\circ$, and (a) $a_2/a_1=1$, $\sin2\beta=0.48$; (b) $a_2/a_1=2$, $\sin2\beta=0.48$; (c) $a_2/a_1=1$, $\sin2\beta=0.64$; (d) $a_2/a_1=1$, $\sin2\beta=0.32$.  The solid line is the ``true'' value of $\gamma$.}}
Our first observation is that the accuracy in $\gamma$ is relatively insensitive to the values of $a_2/a_1$ and $\sin2\beta$.  To illustrate this, we fix $\gamma=110^\circ$, $\Delta=10\%$, $\bar\delta=10^\circ$, and scan over the values of $\sin2\beta$ and $a_2/a_1$ given in Table~\ref{table:params}.  The results are plotted in Fig.~\ref{fig:gamplot1}, in which we see little variation in the widths of the bands.  The insensitivity to $a_2/a_1$ is perhaps a little surprising, since we known that $a_1$ and $a_2$ must be of the same magnitude for the method to work.

A much more significant effect on the uncertainty on $\gamma$ is the strong phase $\delta$.  Although $\delta$ is eliminated in the extraction of $\gamma$, its value affects the shape of the amplitude triangles.  This shape determines, in turn, the sensitivity of the analysis to the experimental error $\Delta$.  Fixing $a_2/a_1=1$, $\sin2\beta=0.48$, and $\gamma=70^\circ$, in Fig.~\ref{fig:gamplot2} we show the result of varying $\Delta$ and $\bar\delta$ over the values in the last two lines of Table~\ref{table:params}.  In Fig.~\ref{fig:gamplot3}, we do the same for $\gamma=110^\circ$.  Note that in every case the uncertainty in $\gamma$ grows significantly if $\bar\delta$ is as large as $60^\circ$.
\FIGURE{\epsfig{file=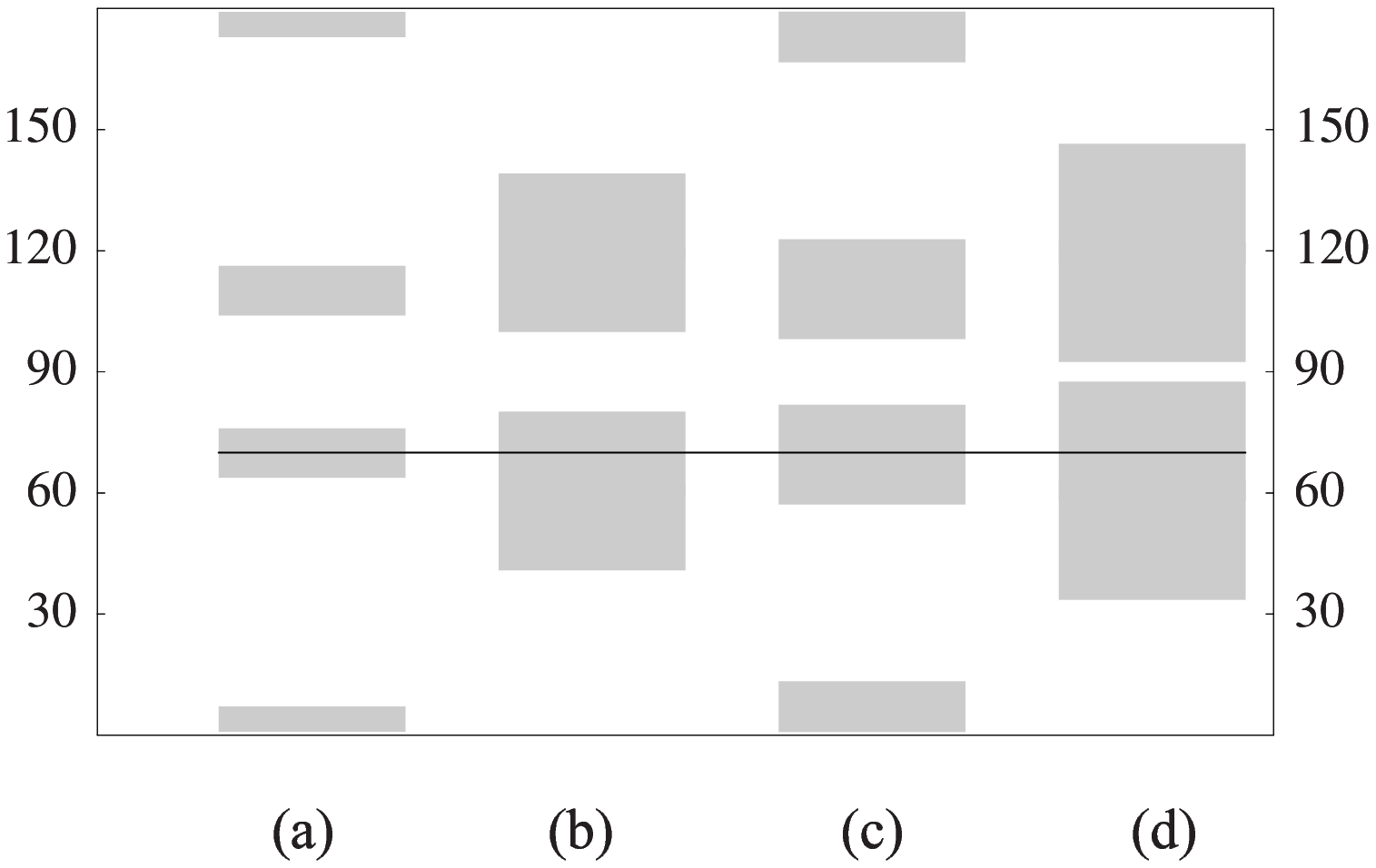,width=12cm}
\caption{\label{fig:gamplot2} Allowed bands for $\gamma$, with input values $\gamma=70^\circ$, $a_2/a_1=1$, $\sin2\beta=0.48$, and (a) $\Delta=10\%$, $\bar\delta=10^\circ$; (b) $\Delta=10\%$, $\bar\delta=60^\circ$; (c) $\Delta=20\%$, $\bar\delta=10^\circ$; (d) $\Delta=20\%$, $\bar\delta=60^\circ$.  The solid line is the ``true'' value of $\gamma$.}}
\FIGURE{\epsfig{file=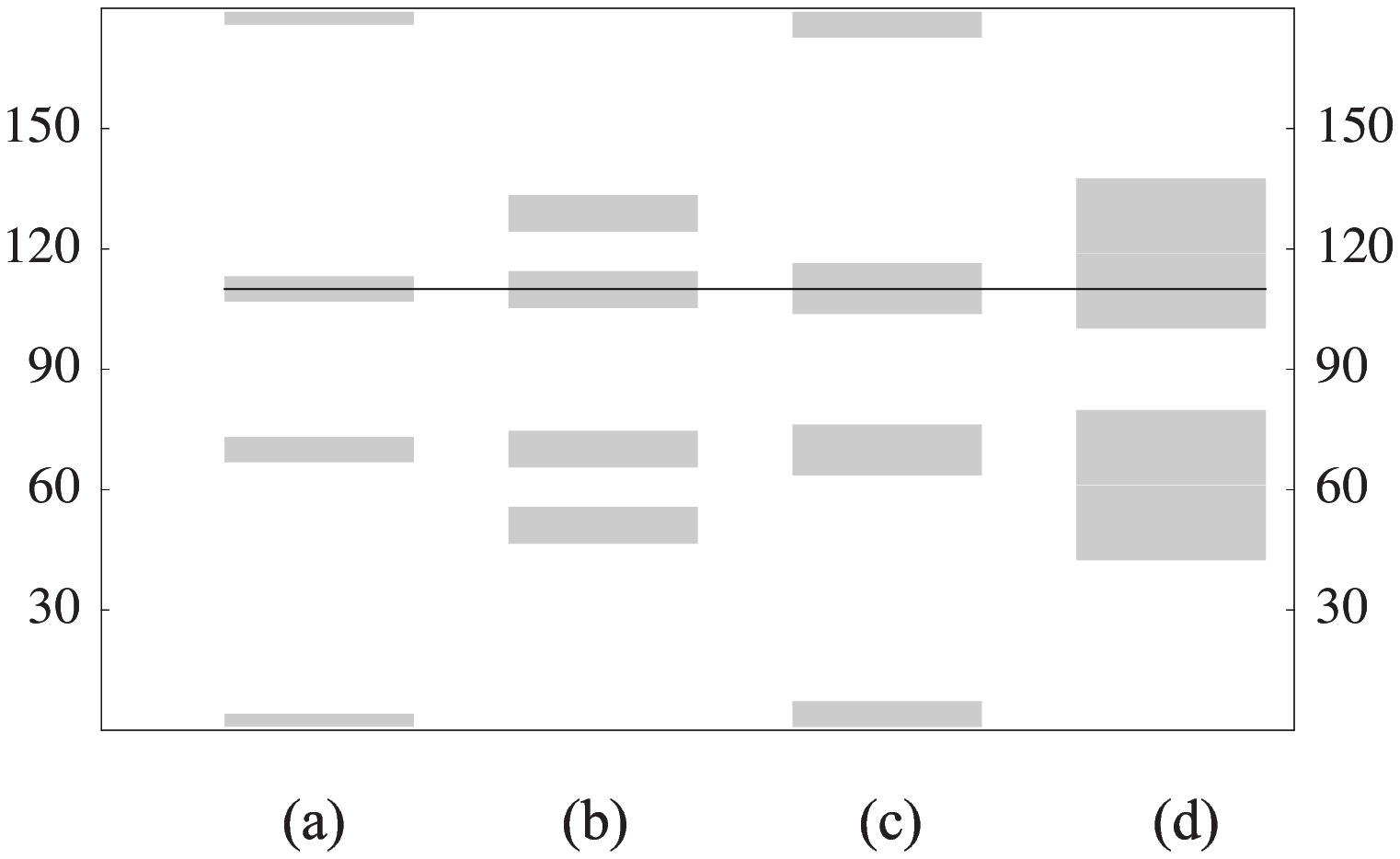,width=12cm}
\caption{\label{fig:gamplot3} The same inputs as in the previous figure, but with $\gamma=110^\circ$.}}

We see that the accuracy with which $\gamma$ can be extracted depends almost as significantly on the values of $\gamma$ and $\bar\delta$ as it does on the precision $\Delta$ of the experimental measurement.  For example, if $\Delta=20\%$, the case $\gamma=110^\circ$ and $\bar\delta=10^\circ$ (plot \ref{fig:gamplot3}c) gives a total allowed region in $\gamma$ of $38^\circ$, while by contrast the case $\gamma=70^\circ$ and $\bar\delta=60^\circ$ (plot~\ref{fig:gamplot2}d) gives an almost useless allowed region of $108^\circ$.  A particularly favorable case, if a 10\% measurement could be made instead, is $\gamma=110^\circ$ and $\bar\delta=10^\circ$ (plot \ref{fig:gamplot3}a), for which the allowed region would be only $19^\circ$.  

If we take the point of view that the ambiguity will be resolved by other measurements, then the  extraction of $\gamma$ by this method appears considerably more significant.  Given our ealier discussion, let us also discard the possibility of large $\bar\delta$.  From this perspective, the $\Delta=10\%$ measurement, with $\gamma=110^\circ$ and $\bar\delta=10^\circ$, would fix $\gamma$ to $\pm3^\circ$.  Even if $\Delta$ were $20\%$, then $\gamma$ could be determined to $\pm5^\circ$ if $\gamma=110^\circ$ (plot \ref{fig:gamplot3}c), or to $\pm12^\circ$ if $\gamma=70^\circ$ (plot \ref{fig:gamplot2}c). Especially given that there are no theoretical uncertainties in this analysis, such an analysis would be quite competitive with any other technique for measuring $\gamma$ which has been proposed.

\section{Conclusions}\label{sec:conclude}

A next generation $B$ Factory operating at the $\Upsilon(4S)$ could yield in a couple of years of running a sample of a million decays of $B$ mesons initially in an eigenstate of CP.  The analysis of these decays would be a brand new method available only to a detector such as Super\babar.  Such a sample will be useful even in the presence of CP violation in $B_d$ mixing, if this effect is well measured independently.  In this paper, we have discussed one particularly significant analysis which would be possible in this era, namely the extraction of $\gamma$ from the decay $B_{\rm CP}\to DK_S$.  The eventual accuracy with which $\gamma$ could be determined depends most significantly on the values of $\gamma$ and an unknown strong phase $\delta$.  Generally, we find that a measurement of the CP tagged branching ratios with a precision of $20\%$ or better could reasonably yield $\gamma$ to within approximately $5^\circ$.  There is an eightfold ambiguity in the extraction which we assume would be resolved by recourse to independent determinations of $\gamma$.

While this analysis is certainly promising, given the experimental uncertainties it probably does not yet rise to the level of a ``flagship'' experiment for Super\babar\ or a successor to Belle.  Yet if a next generation $B$ Factory is eventually built, a large sample of CP tagged $B_d$ decays will be an important component of the program.  Applied to the extraction of $\gamma$ or to other purposes, one can hope that it will provide a useful new avenue for $B$ physics and CKM phenomenology.

\acknowledgments
It is a pleasure to thank K.~J.~Falk for a careful and critical reading of the manuscript.
Support was provided by the U.S.~National Science
Foundation under Grant PHY--9970781 and by the  U.S.~Department of
Energy under Outstanding Junior Investigator Award
DE--FG02--94ER40869.  A.F.~is a Cottrell Scholar of the Research
Corporation.

\end{document}